\documentstyle[preprint,aps]{revtex}
\newcommand{\be}{\begin{equation}} \newcommand{\ee}{\end{equation}}
\newcommand{\bea}{\begin{eqnarray}}\newcommand{\eea}{\end{eqnarray}}

\begin{document}
\draft
\preprint{IP/BBSR/94-14}
\title { Self-Dual Charged Vortices of Finite Energy per Unit Length
in $3+1$ Dimensions}
\author{Pijush K. Ghosh\cite{mail} and Avinash Khare\cite{mail1}}
\address{Institute of Physics, Bhubaneswar-751005, INDIA.}
\maketitle
\begin{abstract}
We obtain both topological as well as nontopological self-dual charged
vortex solutions of finite energy per unit length in a generalized abelian
Higgs model in $3+1$ dimensions. In this model the Bogomol'nyi bound on the
energy per unit length is obtained as a linear combination of the magnetic
flux and the electric charge per unit length.
\end{abstract}
\narrowtext

\newpage

It is well known that the Ginzburg-Landau model of
superconductivity\cite{landau} and also its relativistic generalization,
i.e., the abelian Higgs model admits topologically stable vortex solutions
of finite energy per unit length in $3+1$ dimensions\cite{nil}. These
vortices have received considerable attention in the literature
because of their possible relevance in the context of cosmic strings
as well as superconductivity. These vortices are electrically neutral.
Infact in $1975$
Julia and Zee\cite{julia} showed that unlike the case of dyons in $SO(3)$
Georgi-Glashow model, the abelian Higgs model does not admit charged
generalization. Sometime ago, one of us (AK) with Paul showed\cite{paul}
that in $2+1$ dimensions this Julia-Zee objection can be overcomed and
one can have charged vortices ( solitons to be more precise ) of finite
energy in the abelian Higgs model with Chern-Simons ({\bf CS}) term.
However, to the best of our knowledge, as far as $3+1$ dimensions
are concerned, no one has been able to overcome the Julia-Zee objection
and obtain charged vortex solutions of finite energy per unit length.

The purpose of this letter is to show that the Julia-Zee objection
can be overcomed in $3+1$ dimensions and one can have charged
vortices of finite energy per unit length. We consider a generalized abelian
Higgs model with a dielectric function and a neutral scalar field and show
that such a model admits self-dual toplogical as well as nontopological
charged vortex solutions of finite energy per unit length. Remarkably
enough, the Bogomol'nyi equations\cite{bogo} of our model can be shown
to be essentially identical to the corresponding equations of the pure
{\bf CS} Higgs vortices\cite{hong}. However, unlike in that case, the
Bogomol'nyi bound on the energy per unit length is obtained as a linear
combination of the magnetic flux and the electric charge per unit
length. As a result, unlike in the {\bf CS} case, the nontopological
self-dual charged vortices turn out to be unstable against
decay to the elmentary excitations. Finally  using
the cylindrical
ansatz, we show that the angular momentum and the magnetic moment of
the vortices can also be computed analytically.

Let us consider the following generalized abelian Higgs model
\bea
{\cal L} \ & = & \ -{1 \over 4} G(\mid \phi \mid) F_{\mu \nu} F^{\mu \nu}
+ {1 \over 2} {\mid (\partial_\mu-i e A_\mu)\phi \mid}^2 +
{1 \over 2} G({\mid \phi \mid}) \partial_\mu N \partial^\mu N\nonumber \\
& & - {e^2 \over {8 G({\mid \phi \mid})}} ({\mid \phi \mid}^2-v^2)^2
- {e^2 \over 2} N^2 {\mid \phi \mid}^2
\label{eq1}
\eea
\noindent where $G({\mid \phi \mid})$ is the scalar field dependent
dielectric function while $N$ is a massless neutral scalar filed.
The modification to the Maxwell kinetic energy term can be viewed as an
effective action for a system in a medium described by a suitable
dielectric function. Infact, certain soliton bag models
are described by a Lagrangian where such
a dielectric function
is multiplied with the Maxwell kinetic energy term\cite{soliton}.
Further, in certain
supersymmetric theories
such a non-minimal kinetic term is necessary in order to have a
sensible gauge theory\cite{hull}. Even in the context of vortex solutions,
such non-minimal Maxwell kinetic energy term has been considerd before and
Bogomol'nyi
bounds have been obtained in the case of both neutral\cite{nam}
and charged
{\bf CS} vortices\cite{{nam},{lohe}}. Infact
the Lagrangian (\ref{eq1}) is a special case of a {\bf CS} charged vortex
model
considered in Ref. \cite{nam} ( see their eq. (19) ) when the coefficient
of the {\bf CS} term in that model is put equal to zero.

The field equations that follow from the Lagrangian (\ref{eq1}) are
\be
D_\mu (D^\mu \phi) + {{\partial G({\mid \phi \mid})} \over
{\partial \phi^*}} ( {1 \over 2} F_{\mu \nu} F^{\mu \nu} -
\partial_\mu N \partial^\mu N ) +
2 {{\partial V({\mid \phi \mid})} \over {\partial \phi^*}}=0
\label{eq2}
\ee
\be
\partial_\mu ( G({\mid \phi \mid}) F^{\mu \nu} ) = J^\nu
\label{eq3}
\ee
\be
\partial_\mu (G({\mid \phi \mid}) \partial^\mu N ) =
- e^2 N {\mid \phi \mid}^2
\label{eq4}
\ee
\noindent where the conserved Noether current $J_\mu$ is defined as
\be
J\mu= - {{i e} \over 2} [\phi^* (D_\mu \phi) - \phi (D_\mu \phi)^*]
\label{eq5}
\ee
\noindent The energy momentum tensor $T_{\mu \nu}$ that follows from
Lagrangian (\ref{eq1}) is
\bea
T{\mu \nu} \ & = & \ {1 \over 2} [ (D_\mu \phi) (D_\nu \phi)^* +
(D_\nu \phi) (D_\mu \phi)^* ]\nonumber \\
& & + G({\mid \phi \mid}) ( F_{\mu \alpha}
F^{\alpha} \;_{\nu} + \partial_\mu N \partial_\nu N)
-g_{\mu \nu} {\cal L}
\label{eq6}
\eea
Using the Bogomol'nyi trick, the energy per unit length $E$ can be
written as
\bea
E \ & = & \ {1 \over 2} \int d^2 x [ {\mid (D_1 \pm i D_2) \phi \mid}^2+
{\mid D_0 \phi \pm i e \phi N \mid}^2 +
G({\mid \phi \mid}) ( F_{i 0} \mp \partial_i N )^2\nonumber \\
& & + G({\mid \phi \mid}) \; ( F_{12} \pm {e \over {2 G({\mid \phi \mid})}}
({\mid \phi \mid}^2 - v^2) \; )^2 + G({\mid \phi \mid})
(\partial_0 N)^2]\nonumber \\
& & \pm {{e v^2} \over 2} \Phi \pm \int d^2 x \partial_i ( N
G({\mid \phi \mid}) F_{i0} )\nonumber \\
& & \geq \pm {{e v^2} \over 2} \Phi
\pm \int d^2 x \partial_i (N G({\mid \phi \mid}) F_{i0})
\label{eq7}
\eea
\noindent The bound on the energy is thus saturated when the
following Bogomol'nyi equations hold true
\be
(D_1 \pm i D_2) \phi = 0
\label{eq8}
\ee
\be
D_0 \phi \pm i e \phi N = 0
\label{eq9}
\ee
\be
 F_{i 0} \mp \partial_i N = 0
\label{eq10}
\ee
\be
 F_{12} \pm {e \over {2 G({\mid \phi \mid})}}
({\mid \phi \mid}^2 - v^2) = 0
\label{eq11}
\ee
\be
\partial_0 N =0
\label{eq12}
\ee
\noindent From eqs. (\ref{eq8}) and (\ref{eq11}) one can easily show that
that away from the zeros of $\phi$, it obeys the uncoupled equation
\be
\bigtriangledown^2 ln {\mid \phi \mid}^2 + {e^2 \over G({\mid
\phi \mid})} (v^2 - {\mid \phi \mid}^2) = 0
\label{eq13}
\ee
Further, eqs. (\ref{eq9}), (\ref{eq10}) and (\ref{eq12}) are automatically
satisfied if one considers static solutions and have $N=\pm A_0$.

Apart from the Gauss law equation (i.e., field equation (\ref{eq3}) with
$\nu=0$ ), all other field equations are also automatically satisfied once
eqs. (\ref{eq8}) to (\ref{eq13}) hold true. We now observe that in case the
dielectric function $G({\mid \phi \mid})$ is chosen to be
\be
G({\mid \phi \mid})=g_0 (e {\mid \phi \mid})^{-2}
\label{eq14}
\ee
\noindent then the Gauss law equation ( which is second order in nature )
is consistent with eq. (\ref{eq13}) provided
\be
A_0=\mp {e h_0 } (v^2 - {\mid \phi \mid}^2)
\label{eq15}
\ee
\noindent Here $g_0$ and $h_0$ are arbitrary constants with mass dimension
of $2$ and $-1$ respectively. It is amusing to note that for
$h_0={1 \over 2}$ the
Bogomol'nyi equations (\ref{eq8}), (\ref{eq11}) and (\ref{eq15}) with
$G({\mid \phi \mid})$ as given by (\ref{eq14}) are identical to those of
the pure {\bf CS} Higgs vortices\cite{hong} and hence most of the
results of that model can be taken over in our case.

Let us now address the key point of this letter, i.e., to show that one
has indeed charged vortices with finite energy per unit length. From the
Gauss law eq. (\ref{eq3}) ( with $\nu=0$ ) we find that the charge per
unit length $Q$ is given by
\be
Q=-\int J_0 d^2 x=e^2 \int d^2 x A_0 {\mid \phi \mid}^2
\label{eq16}
\ee
\noindent where use has been made of eq. (\ref{eq5}). In view of eqs.
(\ref{eq11}) and (\ref{eq15}) we then find that the charge $Q$ is nonzero
and given by
\be
Q=-2 h_0 g_0 \Phi
\label{eq17}
\ee
\noindent where the flux $\Phi=\int F_{12} d^2 x$. Thus in case the flux is
quantized then the vortex charge is also automatically quantized.

Finally, using eqs. (\ref{eq11}) to (\ref{eq16}) in eq. (\ref{eq7}) it is
easily shown that in the Bogomol'nyi limit, the energy of the self-dual
vortices is obtained as a linear combination of the magnetic flux and
the electric charge per unit length, i.e.,
\be
E=\pm {{e v^2} \over 2} \Phi \mp e v^2 h_0 Q = \pm {{e v^2} \over 2}
(1+4 h_0^2 g_0) \Phi
\label{eq18}
\ee

The self-dual eqs. (\ref{eq8}), (\ref{eq11}) and (\ref{eq13}) admit both
topological as well as nontopolgical charged vortex solutions which can be
analyzed in detail by following the work of refs. \cite{wang} and
\cite{yang}. In particular, following Wang\cite{wang} it can be
rigorously shown that the toplogical charged vortex solutions to
eq. (\ref{eq13})
exist and are unique satisfying ${\mid \phi \mid} \rightarrow v$ at
spatial infinity. They have quantized flux ($\Phi={{2 \pi} \over e} n$),
charge per unit length ($Q=-2 h_0 g_0 \Phi$) and energy per unit
length ($E={{e v^2} \over 2} (1+4 h_0^2 g_0) \Phi$).
Note that whereas for the neutral vortex the magnetic field is
maximum at the core, for this charged vortex it vanishes at the core
and is maximum in a ring around it.
Further,
the $n$-vortex solution is described by $2 \; n$ continuous parameters
characterizing the position of the $n$ noninteracting vortices.

For the nontopological self-dual charged vortices, ${\mid \phi \mid}
\rightarrow 0$ at spatial infinity. As a result neither flux nor
charge nor energy are quantized and hence these are not the minimum
energy configurations. For these solutions, the energy per unit
charge is given by ( see eq. (\ref{eq18}) )
\be
{E \over {\mid Q \mid}} = {m \over e} ( 2 h_0 \sqrt{g_0}+
{1 \over {2 h_0 \sqrt{g_0}}}) > {m \over e}
\label{eq19}
\ee
\noindent where $m={{e^2 v^2} \over {2 \sqrt{g_0}}}$ is the mass of the
elementary excitation in the theory. Thus unlike the {\bf CS} Higgs
vortices\cite{hong}, these nontopological vortices are not stable against
decay to the elementary excitations.

It is of some interest to consider the $n$-vortex solutions
in the cylindrical ansatz in which case the $n$-vortices are superimposed
on the top of each other. On using the ansatz
\be
\phi=v f(r) e^{- i n \theta}, \; \; \vec{A} = - \hat{e_\theta}
v \lambda {{a(r)-n} \over r}, \; \; A_0=v \lambda g(r)
\label{eq20}
\ee
\noindent where $\lambda={{e v} \over \sqrt{g_0}}$
and $r=e v \lambda \rho$
are dimensionless. The angular momentum of these vortices is
easily computed and one finds that $J_z=-4 \pi v^2 h_0 n^2$ or
$4 \pi v^2 h_0 (\alpha^2-n^2)$ depending on
whether it is topological or nontopolgical vortex respectively. Here
$\alpha= -a(\infty)$ and it can be rigorously shown
that $\alpha \geq n+2$\cite{khare}.
Further one can also analytically calculate the magnetic moment of
these vortices and show that $\mu_z={{2 \pi} \over {e \lambda^2}} (n^2+
{\mid n \mid})$ or $-{{2 \pi} \over {e \lambda^2}} (\alpha+n) (\alpha
-n-1)$ depending on if they
are topologfical or nontopological vortices respectively\cite{khare}.

This paper raises several questions which need to be looked into. Some
of these are
(i) By now several self-dual vortex solutions are known both in
$2+1$ and $3+1$ dimensions and in most cases one finds that
there is an underlying supersymmetry in the problem\cite{susy}. Thus it
may be worthwhile to enquire if the Lagrangian (\ref{eq1}) is also the
bosonic part of an underlying supersymmetric field theory.
(ii) What happens when one couples the charged vortex solutions to
fermions? In the neutral case, it is known that there is an index
theorem\cite{eric}
and that an $n$-vortex has precisely $n$ zero modes which have been
explicitly found\cite{rossi}.
(iii) Can one couple this model to gravity and again obtain charged
vortex solutions in the full theory? More generally, a la neutral
vortex case
are these charged vortices also relevant in the context of early universe?
(iv) Can one obtain this model by dimensionally reducing self-dual Y. M.
equations?
(v) Can one also obtain self-dual charged vortices in the nonrelativistic
theory a la Jackiw-Pi\cite{pi}?
(vi) Can one construct semi-local charged vortex solutions a la
Vachaspati\cite{semi-local}?
(vii) Finally, perhaps the most important question is if these charged
vortices could be experimentally observed in either normal or high- $T_c$
superconductors? In this context, notice that our entire discussion
is also valid in $2+1$ dimensions and our solutions can also be
regarded as charged vortices ( solitons to be more precise ) of finite
charge and energy in $2+1$ dimensions.

We hope to address some of these issues in the near future.

\acknowledgements
It is a pleasure to acknowledge some useful discussions with Charanjit
Aulakh.

\end{document}